\documentclass[preprint]{JHEP}

\usepackage{epsfig}
\usepackage{tabularx}
\usepackage{graphicx}
\usepackage{amssymb,amsfonts,amsmath}

\title{Non-thermal leptogenesis and gravitino problem in inflaton decay}

\author{Grigoris Panotopoulos\\
ASC, Department of Physics LMU,\\
Theresienstr. 37, 80333 Munich, Germany\\
{\tt E-mail: Grigoris.Panotopoulos@physik.lmu.de}}

\abstract{In the present work we discuss baryon asymmetry in the non-thermal leptogenesis scenario and gravitino cosmology for an unstable gravitino with inflaton decay. We take into account two production mechanisms for gravitino, namely thermal production and inflaton decay. We wish to show in plots the allowed parameter space so that the BBN constraint and the requirement for the right baryon asymmetry are satisfied at the same time. However our analysis shows that it is impossible to achieve both goals simultaneously.}

\begin{document}

\maketitle 

\setcounter{equation}{0}

\section{Introduction}

Supersymmetry~\cite{Martin:1997ns} is the most common way beyond the standard model of particle physics. It is a well-motivated theoretical idea according to which to every known particle corresponds a new particle, its supersymmetric partner. Supersymmetry solves the hierarchy problem, provides a popular candidate for cold dark matter and is the basis for the superstring theory. Furthermore, in a specific realization of supersymmetry, namely the Minimal Supersymmetric Standard Model, experimental data support the unification of the gauge coupling constants at $M_{GUT} \simeq 2 \times 10^{16}$~GeV. Exact supersymmetry predicts that superpartners must have equal masses. However supersymmetry must be broken and superpartners are heavier than ordinary particles. Graviton is massless, but its supersymmetric particle, the gravitino, has a mass that depends on how the supersymmetry is broken. In gravity-mediated supersymmetry breaking the gravitino mass is $m_{3/2} \sim $ 100~GeV-1~TeV and the gravitino is unstable with a lifetime $\tau \sim \frac{M_{p}^2}{m_{3/2}^3}$, which is much larger than the time at Big-Bang Nucleosynthesis, $t_{nuc} \sim $1~sec. The gravitino can be dangerous for cosmology, unless the reheating temperature $T_R$ after inflation is restricted in a certain way. If gravitinos are produced from the thermal bath, then it turns out that the abundance of gravitinos is proportional to the reheating temperature and that the so-called ``gravitino problem''~\cite{Ellis:1982yb} is avoided if $T_R \leq (10^6-10^7)$~GeV~\cite{Cyburt:2002uv}, for $m_{3/2}=100$~GeV for example.

Inflation~\cite{inflation} solves the problems of standard Hot Big-Bang cosmology and at the same time produces the cosmological fluctuations for the structure formation we observe today. It is widely believed that some time in the very early stages during its history, the universe undergoes a rapid acceleration while its energy density is dominated by a scalar field, called the inflaton, with a self interaction potential $V(\phi)$. The inflaton is initially displaced from the minimum of its potential and slowly rolls towards the minimum. Then the inflaton decays to ordinary particles that thermalize and the universe reenters into the standard FRW phase. Since supersymmetry is a popular way beyond the standard model it is natural to discuss inflation in the framework of supergravity. It has been recently pointed out that gravitinos are generically produced by inflaton decay~\cite{Kawasaki:2006gs}. This is therefore another gravitino production mechanism which occurs in most inflationary models in supergravity. It would be interesting to study the effect of inflaton decay on the conventional gravitino cosmology.

One of the theoretical challenges for modern cosmology is to explain the baryon asymmetry in the universe (BAU). From Big-Bang Nucleosynthesis (BBN)~\cite{Steigman:2005ys} and WMAP~\cite{Bennett:2003bz} we know that the ratio of the baryon density $n_B$ over the photon density $n_{\gamma}$ is a very small number
\begin{equation}
\eta \equiv \frac{n_B}{n_{\gamma}}=(6.1 \pm 0.3) \times 10^{-10}
\end{equation}
Today the most popular way of explaining the BAU is baryogenesis through leptogenesis~\cite{Fukugita:1986hr}. Initially a lepton asymmetry is generated and then it is partially converted to baryon asymmetry via ``sphaleron'' effects~\cite{Kuzmin:1985mm}. Leptogenesis can be thermal or non-thermal. Generally thermal leptogenesis demands high reheating temperature, $T_R \geq (2-3) \times 10^9$~GeV~\cite{Buchmuller:2002rq}, which can be problematic because of the gravitino problem. On the other hand, non-thermal leptogenesis can produce the right lepton asymmetry even if the reheating temperature is lower~\cite{Asaka:1999yd} compared to that of thermal leptogenesis.

Before proceeding, let us first summarize our work here. We consider a heavy, unstable gravitino with a mass in the $(0.1-1)$~TeV range. We compute the total gravitino yield taking into account both thermal and non-thermal production for gravitino and then impose the BBN constraint. Furthermore, we require the right baryon asymmetry in the framework of non-thermal leptogenesis. Then we determine the allowed parameter space in order that both requirements (right asymmetry and avoiding the gravitino overproduction problem) are satisfied simultaneously.

Our work is organized as follows: After this introduction we present the basic formulae in section 2 and then we obtain the allowed parameter space in the third section. Finally we conclude in the last section.

\section{Theoretical framework}

\subsection{Gravitino cosmology}

We consider the popular scenario of gravity mediated supersymmetry breaking and take the gravitino mass to be in the range $100~GeV \leq m_{3/2} \leq 1~TeV$. The gravitino is considered to be unstable and its lifetime is given by~\cite{Nakamura:2006uc}
\begin{equation}
\tau_{3/2}^{-1}=\Gamma_{3/2}=\frac{193}{384 \pi} \: \frac{m_{3/2}^3}{M_p^2}
\end{equation}
where $M_p=2.4 \times 10^{18}$~GeV is the reduced Planck mass. We can check that the gravitino lifetime $\tau_{3/2} > 10^4~sec$ and therefore it decays after BBN starts. Energetic particles produced by the gravitino decay may dissociate the background nuclei and significantly affect the primordial abundances of the light elements. If such processes occur with sizable rates, the predictions of the standard BBN scenario are altered and the success of the primordial nucleosynthesis is spoiled. BBN constraints on cosmological scenarios with exotic long-lived particles predicted by physics beyond the Standard Model have been studied in~\cite{constraints}. These studies set an upper bound on the gravitino yield, $Y_{3/2} < \zeta_{max}$, where $Y_{3/2}=n_{3/2}/s$, $s=h_* (2 \pi^2/45) T^3$ is the entropy density and $h_*$ counts the effective number of relativistic degrees of freedom. Here we shall use the values cited in~\cite{constraints}b, according to which $\zeta_{max}=3 \times 10^{-16}$ for $m_{3/2}=100$~GeV and $\zeta_{max}=4.5 \times 10^{-17}$ for $m_{3/2}=1$~TeV.

The total gravitino yield has two contributions, namely a thermal and a non-thermal one
\begin{equation}
Y_{3/2}=Y_{3/2}^{TP}+Y_{3/2}^{NTP}
\end{equation}
The contribution from the thermal production has been computed in~\cite{Bolz:2000fu,Pradler:2006qh,Rychkov:2007uq}. In~\cite{Bolz:2000fu} the gravitino production was computed in leading order in the gauge coupling $g_3$, in~\cite{Pradler:2006qh} the thermal rate was computed in leading order in all Standard Model gauge couplings $g_Y, g_2, g_3$, and in~\cite{Rychkov:2007uq} new effects were taken into account, namely: a) gravitino production via gluon $\rightarrow$ gluino $+$ gravitino and other decays, apart from the previously considered $2 \rightarrow 2$ gauge scatterings, b) the effect of the top Yukawa coupling, and c) a proper treatment of the reheating process. Here we shall use the result of~\cite{Bolz:2000fu} since the corrections of~\cite{Pradler:2006qh,Rychkov:2007uq} do not alter our conclusions. Therefore the thermal gravitino production is given by
\begin{equation}
Y_{3/2}^{TP}=10^{-12} \left ( 1+\frac{m_{\tilde{g}}^2}{3 m_{3/2}^2}  \right) \: \left ( \frac{T_R}{10^{10}~GeV} \right )
\end{equation}
with $m_{\tilde{g}}$ the gluino mass taken to be $m_{\tilde{g}}=300$~GeV,
The second contribution from inflaton decay is given by
\begin{equation}
Y_{3/2}=\frac{3}{2} \: \frac{\Gamma_{3/2}}{\Gamma_\phi} \: \frac{T_R}{m_\phi}
\end{equation}
where $T_R$ is the reheating temperature after inflation, $m_\phi$ is the inflaton mass, $\Gamma_\phi$ is the total decay rate of the inflaton, and $\Gamma_{3/2}$ is the inflaton decay rate to a pair of gravitinos. The total decay rate of the inflaton is given in terms of the reheating temperature
\begin{equation}
\Gamma_\phi = \left(\frac{\pi^2 g_*}{10} \right)^{1/2} \frac{T_R^2}{M_p}
\end{equation}
where $g_*$ counts the relativistic degrees of freedom.
Assuming a non-vanishing expectation value for the inflaton, its decay rate to gravitinos is given by
\begin{equation}
\Gamma_{3/2}=\frac{10^{-4}}{32 \pi} \: \left( \frac{\langle \phi \rangle}{M_p} \right)^2 \: \frac{m_\phi^3}{M_p^2}
\end{equation}
Combining everything together we obtain for the non-thermal contribution~\cite{Takahashi:2007tz}
\begin{equation}
Y_{3/2}^{NTP}=7 \times 10^{-15} \left( \frac{g_*}{200} \right)^{-1/2} \left( \frac{\langle \phi \rangle}{10^{15}~GeV} \right)^2 \left( \frac{m_\phi}{10^{12}~GeV} \right)^2 \left( \frac{10^{6}~GeV}{T_R} \right)
\end{equation}
where $\langle \phi \rangle$ is the inflaton expectation value.

\subsection{Baryon asymmetry in non-thermal leptogenesis}

In the non-thermal leptogenesis scenario~\cite{nonthermal} the heavy neutrinos are produced through the direct non-thermal decay of the inflaton. We start by introducing three heavy right-handed neutrinos (one for each family) $N_i, i=1,2,3$ with
masses $M_{1}, M_{2}, M_{3}$. Supergravity effects enable the inflaton to decay into all matter fields once the inflaton acquires a non-vanishing expectation value. We assume that after the slow-roll phase of inflation the inflaton decays predominantly into the lightest of the heavy neutrinos.  The inflaton decay into $N_1$ is given by~\cite{Endo:2006qk}
\begin{equation}
\Gamma_N \equiv \Gamma(\phi \rightarrow N_{1} N_{1})=\frac{1}{16 \pi} \frac{m_\phi M_1^2}{M_p^2} \left( \frac{\langle \phi \rangle}{M_p} \right)^2 \sqrt{1-\frac{4 M_1^2}{m_\phi^2}}
\end{equation}
Note that one does not have to introduce any direct couplings of the inflaton with the right-handed neutrinos to induce the decay. The decay proceeds as long as the inflaton acquires a nonzero VEV, and it is kinematically allowed provided that
\begin{equation}
m_\phi > 2M_1
\end{equation}
At this point we should mention that we have neglected the inflaton decay into a pair of right-handed neutrinos through direct Yukawa couplings considered in the usual leptogenesis scenario
\begin{equation}
\Gamma_{N}^{(dir)}=\frac{\lambda^2 m_{\phi}}{4 \pi}
\end{equation}
where $\lambda$ is the Yukawa coupling. This is possible since the Yukawa coupling could be tiny. For example, if $M_1 \sim 10^{11}~GeV$ and $\langle \phi \rangle \sim M_p$ then $\Gamma_{dir} \ll \Gamma_N$ provided that $\lambda \ll 10^{-8}$.

Any lepton asymmetry $Y_{L} \equiv n_{L}/s$ produced before the electroweak phase transition is partially converted into a baryon asymmetry $Y_{B} \equiv n_{B}/s$ via sphaleron effects~\cite{Kuzmin:1985mm}. The resulting $Y_B$ is
\begin{equation}
Y_{B}=a \: Y_{L}
\end{equation}
with the fraction $a$ computed to be $C=-8/23$ in the MSSM~\cite{Harvey:1990qw}. The lepton asymmetry, in turn, is generated by the $CP$-violating  out-of-equilibrium decays of the heavy neutrinos\begin{equation}
N \rightarrow l H_{u}^{*}, \quad N \rightarrow \bar{l} H_{u}
\end{equation}
For convenience we parameterize the $CP$ asymmetry  in the neutrino decays  $\epsilon$ in the form
\begin{equation}
\epsilon=\epsilon^{max} \: \textrm{sin} \delta
\end{equation}
where $\delta$ is an effective leptogenesis phase and $\epsilon^{max}$ is the maximum asymmetry which is given by~\cite{Davidson:2002qv}
\begin{equation}
\epsilon^{max}=\frac{3}{8 \pi} \: \frac{M_1 \sqrt{\Delta m_{atm}^2}}{v^2 sin^2 \beta}
\end{equation}
with $v=174$~GeV the electroweak scale, $tan \beta$ the ratio of the vevs of the two Higgs doublets of the MSSM and
$\Delta m_{atm}^2=2.6 \times 10^{-3}~\textrm{eV}^2$ the mass squared difference measured in atmospheric neutrino oscillation experiments. For simplicity we shall take $sin \beta \sim 1$ (large $tan \beta$ regime), in which case the maximum $CP$ asymmetry is given by
\begin{equation}
\epsilon^{max}=2 \times 10^{-10} \: \left( \frac{M_1}{10^{6}~\textrm{GeV}} \right )
\end{equation}

In the framework of non-thermal leptogenesis the lepton asymmetry is given by~\cite{Asaka:1999yd}
\begin{equation}
Y_L=\frac{3}{2} \: BR(\phi \rightarrow N_1 N_1) \: \frac{T_R}{m_\phi} \: \epsilon
\end{equation}
where $BR(\phi \rightarrow N_1 N_1)=\Gamma_N/\Gamma_\phi$ is the branching ratio for the decay of the inflaton to the lightest heavy right-handed neutrino. The final result for the baryon asymmetry is given by~\cite{Takahashi:2007tz}
\begin{equation} \label{BA}
Y_B=10^{-9} \left( \frac{g_*}{200} \right)^{-1/2} \left( \frac{\langle \phi \rangle}{10^{16}~GeV} \right)^2 \left( \frac{M_1}{10^{13}~GeV} \right)^3 \left( \frac{T_R}{10^6~GeV}\right)^{-1} |\textrm{sin} \delta|
\end{equation}

\section{Analysis and the allowed parameter space}

In the previous section we saw that the gravitino yield from thermal production is proportional to the reheating temperature
\begin{eqnarray}
Y_{3/2}^{TP} & = & Y_1 T_R \\
Y_1 & = & \left ( 1+\frac{m_{\tilde{g}}^2}{3 m_{3/2}^2}  \right) \: \left ( \frac{10^{-22}}{GeV} \right )
\end{eqnarray}
while the second contribution from inflaton decay is inversely proportional to the reheating temperature
\begin{eqnarray}
Y_{3/2}^{NTP} & = & \frac{Y_2}{T_R} \\
Y_2 & = & 7 \times 10^{-15} \left( \frac{g_*}{200} \right)^{-1/2} \left( \frac{\langle \phi \rangle}{10^{15}~GeV} \right)^2 \left( \frac{m_\phi}{10^{12}~GeV} \right)^2 10^{6}~GeV
\end{eqnarray}
The BBN constraint yields
\begin{equation}
Y_1 T_R + \frac{Y_2}{T_R} < \zeta_{max}
\end{equation}
or
\begin{equation} \label{quadratic}
Y_1 T_R^2-\zeta_{max} T_R+Y_2 < 0
\end{equation}
First of all the condition
\begin{equation}
\Delta \equiv \zeta_{max}^2-4Y_1Y_2 > 0
\end{equation}
should hold because otherwise the BBN cannot be satisfied. Substituting the expressions for $Y_1, Y_2$ we obtain the first important condition for our analysis
\begin{equation}
m_\phi < \frac{\zeta_{max}}{C \langle \phi \rangle}
\end{equation}
where $C$ is given by
\begin{equation}
C=\sqrt{2.8} \times 10^{-15} \left( \frac{g_*}{200} \right)^{-1/4} \left ( 1+\frac{m_{\tilde{g}}^2}{3 m_{3/2}^2}  \right)^{1/2} \left( \frac{10^{-27}}{GeV^2} \right)
\end{equation}
Provided that the BBN constraint is satisfied, we obtain a lower and an upper bound for the reheating temperature
\begin{equation}
\rho_- < T_R < \rho_+
\end{equation}
where $\rho_-$ and $\rho_+$ are the two positive roots of the quadratic expression in (\ref{quadratic})
\begin{equation}
\rho_{\pm}=\frac{\zeta_{max} \pm \sqrt{\Delta}}{2Y_1}
\end{equation}
We can understand this as follows. For large reheating temperature the thermal production is the dominant contribution in the gravitino yield and the requirement $Y_{3/2} < \zeta_{max}$ leads to an upper bound for the reheating temperature. On the other hand, for low reheating temperature the contribution from inflaton decay is the dominant one and the BBN constraint leads to a lower bound for the reheating temperature.

We go on with the requirement for the right baryon asymmetry. In the expression for $Y_B$, equation (\ref{BA}), we take $Y_B \simeq 8.5 \times 10^{-11}$, while the effective leptogenesis phase $|\textrm{sin} \delta|$ and the heavy neutrino mass $M_1$ can be computed in a certain model for neutrino masses. In order to keep the discussion model independent we use the fact that $|\textrm{sin} \delta| \leq 1$, which leads to another condition for the reheating temperature
\begin{equation}
T_R \leq T_{max}
\end{equation}
where $T_{max}$ is given by
\begin{equation}
T_{max}=\frac{1}{Y_B} 10^{-9} \left( \frac{g_*}{200} \right)^{-1/2} \left( \frac{\langle \phi \rangle}{10^{16}~GeV} \right)^{2} \left( \frac{M_1}{10^{13}~GeV} \right)^{3} 10^6~GeV
\end{equation}
Had we considered the usual scenario with inflaton decay into right-handed neutrinos through direct Yukawa couplings, then the expression for $T_{max}$ would have been
\begin{equation}
T_{max}^{(dir)}=\frac{3 |a| \lambda^2 M_p \epsilon^{max}}{16 \pi^2 Y_B} \: \left ( \frac{10}{g_*} \right )^{1/2}
\end{equation}
Up to now we have derived two conditions for the reheating temperature
\begin{equation}
\rho_- < T_R < \rho_+
\end{equation}
and
\begin{equation}
T_R \leq T_{max}
\end{equation}
and we wish to obtain the allowed parameter space in the ($\langle \phi \rangle, m_\phi$) plane so that both conditions can be satisfied at the same time. It is obvious that $T_{max}$ cannot be lower than $\rho_-$. Therefore we require
\begin{equation}
T_{max} > \rho_-
\end{equation}
This condition can be written in the form
\begin{equation}
\sqrt{\Delta} > \zeta_{max}-2Y_1 T_{max}
\end{equation}
where we have used the expression for $\rho_-$. The easiest way to satisfy this is to demand
\begin{equation}
\zeta_{max}-2Y_1 T_{max} < 0
\end{equation}
which yields
\begin{equation}
\langle \phi \rangle > \sqrt{5 Y_B \zeta_{max}} \left( \frac{g_*}{200} \right)^{1/4} \left( \frac{M_1}{10^{13}~GeV} \right)^{-3/2} \frac{10^{28}~GeV}{\left ( 1+\frac{m_{\tilde{g}}^2}{3 m_{3/2}^2}  \right)^{1/2}}
\end{equation}
Therefore, in the ($\langle \phi \rangle, m_\phi$) plane the allowed parameter space is the area enclosed by the bounds
\begin{eqnarray}
m_\phi & > & 2 M_1 \\
\langle \phi \rangle & > & \langle \phi \rangle_{max} \\
m_\phi & < & \frac{\zeta_{max}}{C \langle \phi \rangle}
\end{eqnarray}
Had we considered the usual scenario with inflaton decay into right-handed neutrinos through direct Yukawa couplings, then in this case we would have found (for $M_1=10^{11}~GeV$)
\begin{equation}
\lambda > 2.2 \times 10^{-8}
\end{equation}
We can instead assume that
\begin{equation}
\zeta_{max}-2Y_1 T_{max} > 0
\end{equation}
which yields
\begin{equation}
\langle \phi \rangle < \sqrt{5 Y_B \zeta_{max}} \left( \frac{g_*}{200} \right)^{1/4} \left( \frac{M_1}{10^{13}~GeV} \right)^{-3/2} \frac{10^{28}~GeV}{\left ( 1+\frac{m_{\tilde{g}}^2}{3 m_{3/2}^2}  \right)^{1/2}}
\end{equation}
To avoid writing long expressions we set
\begin{eqnarray}
T_{max} & = & A \langle \phi \rangle^2 \\
Y_2 & = & B m_{\phi}^2 \langle \phi \rangle^2
\end{eqnarray}
where
\begin{eqnarray}
A & \simeq & 10^{-64} M_1^3~GeV^{-4} \\
B & \simeq & 10^{-62}~GeV^{-3}
\end{eqnarray}
Then we can easily show that that the condition $T_{max} > \rho_{-}$ can be written as follows
\begin{equation}
\frac{\langle \phi \rangle^2}{\frac{\zeta_{max}}{A Y_1}}+\frac{m_{\phi}^2}{\frac{A \zeta_{max}}{B}} < 1
\end{equation}
which is the interior of an ellipsis with axes $a^2=\frac{\zeta_{max}}{A Y_1}, b^2=\frac{A \zeta_{max}}{B}$. At this point we recall the kinematical constraint $m_{\phi} > 2M_1$. The two conditions
\begin{eqnarray}
m_{\phi} & > & 2M_1 \\
1 & > & \frac{\langle \phi \rangle^2}{\frac{\zeta_{max}}{A Y_1}}+\frac{m_{\phi}^2}{\frac{A \zeta_{max}}{B}}
\end{eqnarray}
can only be satisfied at the same time provided that
\begin{equation}
\sqrt{\frac{A \zeta_{max}}{B}} > 2M_1
\end{equation}
which yields to an unacceptable bound for the right-handed neutrino mass
\begin{equation}
M_1 > 10^{18}~GeV
\end{equation}
Therefore this second possibility of satisfying the condition $T_{max} > \rho_{-}$ is excluded and we focus on the first case. We have concluded graphically that for a gravitino mass in the range $m_{3/2}=100~GeV-1~TeV$ and for a right-handed neutrino mass in the range $M_1 \sim (10^{10}-10^{15})~GeV$ there is never an allowed parameter space. As an example, we show in the figure below the two-dimensional parameter space for $m_{3/2}=100~GeV$ and $M_1=5 \times 10^{14}~GeV$.

Had we considered the usual scenario with inflaton decay into right-handed neutrinos through direct Yukawa couplings, then in this case we would have found the condition
\begin{equation}
m_{\phi} \langle \phi \rangle < \tilde{C}
\end{equation}
where the new constant $\tilde{C}$ is given by
\begin{equation}
\tilde{C}=\sqrt{\frac{T_{max}^{(dir)} (\zeta_{max}-Y_1 T_{max}^{(dir)})}{B}}
\end{equation}
and depends on the gravitino mass $m_{3/2}$, the Yukawa coupling $\lambda$, and the right-handed neutrino mass $M_1$. On the contrary, the constant that appears in the corresponding condition (3.8) depends on the gravitino mass only. For $m_{3/2}=100~GeV$, $M_1=10^{11}~GeV$, and $\lambda=10^{-8}$ we find the values
\begin{eqnarray}
\tilde{C} & = & 4.6 \times 10^{25}~GeV^2 \\
\frac{\zeta_{max}}{C} & = & 9.3 \times 10^{25}~GeV^2
\end{eqnarray}
while for lower values of $\lambda$ the constant $\tilde{C}$ is even smaller and $\zeta_{max}/C$ remains the same. Therefore, there are two constraints only, namely
\begin{eqnarray}
m_{\phi} & > & 2M_1 \\
m_{\phi} \langle \phi \rangle & < & \frac{\zeta_{max}}{C}
\end{eqnarray}
or
\begin{eqnarray}
m_{\phi} & > & 2M_1 \\
m_{\phi} \langle \phi \rangle & < & \tilde{C}
\end{eqnarray}
and there is always an allowed parameter space, since now the third condition requiring that $\langle \phi \rangle > \langle \phi \rangle_{min}$ is missing.

\section{Conclusions}

In the present work we have discussed the gravitino overproduction problem and baryon asymmetry in the non-thermal leptogenesis scenario with inflaton decay. We have considered inflationary models in supergravity and a heavy, unstable gravitino which potentially can be dangerous for cosmology. We have taken two gravitino production mechanisms into account, namely thermal production from the thermal bath and non-thermal one from inflaton decay. We have imposed the BBN constraint on the total gravitino yield and we have required for the right baryon asymmetry via non-thermal leptogenesis. For natural values of the gravitino mass and the right-handed neutrino mass we have obtained plots in which the allowed parameter space in principle can be shown. However, according to our results it is not possible to satisfy the BBN constraint and produce the right baryon asymmetry at the same time.

\section*{Acknowlegements}

We would like to thank the anonymous reviewer for his/her valuable comments and suggestions that greatly improved the quality of the article. This work was supported by project "Particle Cosmology".

\newpage

\begin{figure}
\centerline{\epsfig{figure=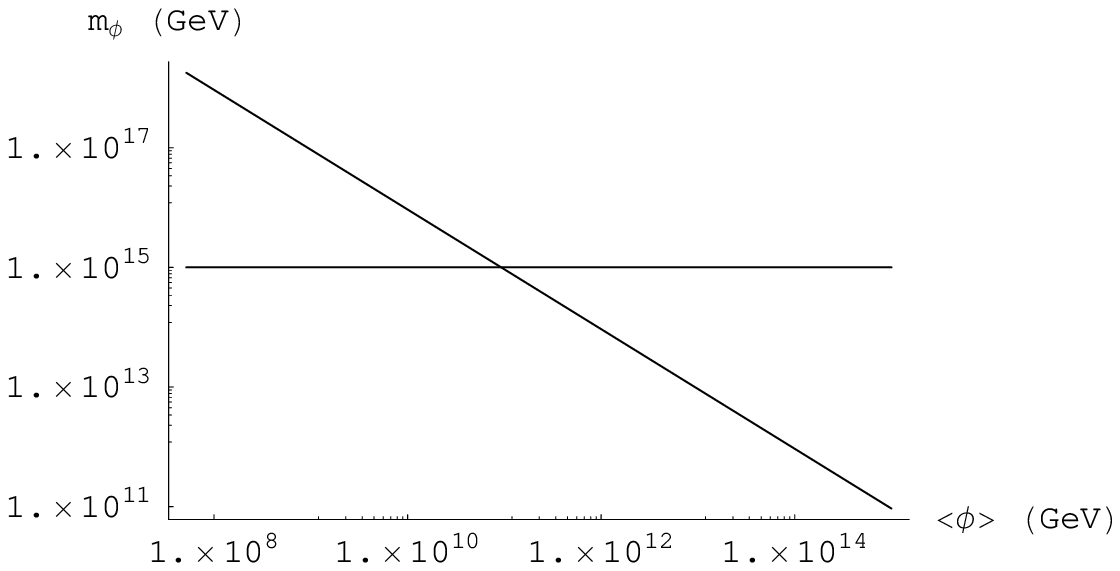,height=8cm,angle=0}}
\caption{Parameter space in logarithmic scale for $m_{3/2}=100~GeV$ and $M_1= 5 \times 10^{14}~GeV$. The two lines correspond to the conditions that $m_{\phi}>2M_1$ and $m_\phi  <  \frac{\zeta_{max}}{C \langle \phi \rangle}$. There is no allowed space, since according to the third condition, $\langle \phi \rangle > 5.22 \times 10^{12}~GeV$.}
\end{figure}

\end{document}